# Too Early? On the Apparent Conflict of Astrobiology and Cosmology

**Milan M. Ćirković**
*Astronomical Observatory of Belgrade*
*Volgina 7, 11160 Belgrade*
*Serbia and Montenegro*
*e-mail:* mcirkovic@aob.bg.ac.yu

**Abstract.** An interesting consequence of the modern cosmological paradigm is the spatial infinity of the universe. When coupled with naturalistic understanding of the origin of life and intelligence, which follows the basic tenets of astrobiology, and with some fairly incontroversial assumptions in the theory of observation selection effects, this infinity leads, as Ken Olum has recently shown, to a paradoxical conclusion. Olum's paradox is related, to the famous Fermi's paradox in astrobiology and SETI studies. We, hereby, present an evolutionary argument countering the apparent inconsistency, and show how, in the framework of a simplified model, deeper picture of the coupling between histories of intelligent/technological civilizations and astrophysical evolution of the Galaxy, can be achieved. This strategy has consequences of importance for both astrobiological studies and philosophy.

## 1. The problem

We are lucky enough to live in an epoch of great progress in the nascent discipline of astrobiology, which deals with three canonical questions: How does life begin and develop? Does life exist elsewhere in the universe? What is the future of life on Earth and in space? A host of important discoveries has been made during the last decade or so, the most important certainly being a discovery of a large number of extrasolar planets; the existence of many extremophile organisms possibly comprising "deep hot biosphere" of Thomas Gold; the discovery of subsurface water on Mars and the huge ocean on Europa, and possibly also Ganymede and Callisto; the unequivocal discovery of amino-acids and other complex organic compounds in meteorites; modelling organic chemistry in Titan's atmosphere; the quantitative treatment of the Galactic habitable zone; the development of a new generation of panspermia theories, spurred by experimental verification that even terrestrial microorganisms easily survive conditions of an asteroidal or a cometary impact; progress in methodology of SETI studies, etc. (for recent beautiful reviews see Des Marais and Walter 1999; Darling 2001; Grinspoon 2003). However, the epistemological and methodological basis of astrobiological and SETI studies presents us with a hornet's nest of issues which have not been, with few exceptions, tackled in the literature so far. It is not surprising, therefore, that seemingly paradoxical situations and conclusions arise from time to time, as is usual in young scientific fields.

For instance, in assessing the importance and future ramification of these discoveries, we obviously take into account our properties as intelligent observers, as well as physical, chemical and other pre-conditions necessary for our existence. The latter are topics of the so-called anthropic reasoning, the subject of much debate and controversy in cosmology, fundamental physics, and philosophy of science. In recent years it became clear that anthropic principle(s) can be most fruitfully construed as *observation selection effects* (Bostrom 2002). This is a straightforward continuation of the Copernican worldview, which emphasizes a non-special character of our cosmic habitat, and which has so immensely contributed to our scientific understanding.

In a recent elegant and thought-provoking paper, Olum (2004) argues that "a straightforward application of anthropic reasoning and reasonable assumptions about the capabilities of other civilizations predicts that we should be part of a large civilization spanning our galaxy." Starting from the assumption of an infinite universe (following from the inflationary paradigm), Olum conjectures that there are civilizations much larger (that is, consisting of much greater number of observers, say $10^{19}$) than ours (about $10^{10}$ observers). Now, even if 90% of all existing civilizations are small ones similar to our own, anthropic reasoning suggests that the overwhelming probabilistic prediction is that we live in a large civilization. Since this prediction is spectacularly unsuccessful on empirical grounds; with a probability of such failure being about $10^{-8}$, something is clearly wrong here.

We shall refer to the alleged incompatibility of anthropic reasoning with observations as "Olum's problem". In a less refined manner, it has been foreseen by J. Richard Gott (in a founding paper in *Nature* on the "Doomsday Argument"; Gott 1993):

> In the limit where (cosmology permitting) a supercivilization is able to accumulate an infinite amount of elapsed conscious time and an infinite number of intelligent observers..., the fraction of ordinary civilizations such as ours that will develop into such a supercivilization must go to zero so that the set of observers born on the original home planet is not an infinitesimal minority of all intelligent observers.

However, the cosmological background has not been so clear at the time of Gott's paper, so a finite universe still was a rather widely accepted option. Recent results (in particular those of WMAP; Bennett et al. 2003; Schwarz and Terrero-Escalante 2004) have strongly confirmed predictions of the inflationary paradigm, which is generically eternal and faces us with strong form of Olum's problem. The same applies even more forcefully to cosmologies following from M-theory (e.g., Billyard et al. 2000). In all these cases, we face an infinite universe with all philosophical conundrums infinities have traditionally presented us with; moreover, new problems, such as Olum's, are bound to arise.

Olum proceeds to give several possible solutions (conveniently labeled as subsections of his section 2) to the problem why we don't find ourselves members of a large civilization:

2.1. anthropic reasoning doesn't work;
2.2. anthropic reasoning should use civilizations instead of individuals;
2.3. one should consider observers who live at any time;
2.4. selection biases;
2.5. infinitesimally few civilizations become large;
2.6. the universe is not infinitely large;
2.7. colonization of the Galaxy is impossible;
2.8. we are a "lost colony" of a large civilization;
2.9. the idea of individual will be different in the future;

admitting that none of them are very likely, but maybe

2.10. some combination of them might work to alleviate the problem.

In this note, we advance a proposal for the general solution which escapes Olum's analysis, although it has some connection with several of his proposals. Let us start with three auxiliary comments.

## 2. Comments

(I) <u>Fermi's paradox as a boundary condition.</u> Olum's problem, as formulated above, can be seen as a generalised version of the well-known Fermi's paradox in astrobiology: *why aren't extraterrestrials here, when they have had so much time to come to us?* It is easy to see how the two are related: suppose that you define a civilization(*) as "all those intelligent observers who are in contact at present". At present, humans are only in contact with other humans. Thus, it makes sense to ask the problematic question: how do we find ourselves belonging to a small civilization(*)? If we are to be contacted by a large alien civilization, then our civilization(*) will be suddenly enlarged, so the corresponding Bayesian probability will increase. A large amount of literature exists on Fermi's paradox (e.g., Brin 1983; Webb 2002); some of the solutions proposed have generalisations applicable to this case, most notably the idea, believed by scientists such as Sagan or von Hoerner, that most civilizations destroy themselves upon discovering nuclear power or a similar risky technology. While we certainly should *not* be surprised that our civilization has managed (so far!) to escape destruction through nuclear war/winter, it is obvious that there is nothing non-exclusive about this: it is perfectly plausible that some civilizations escape these suicidal temptations. Another approach is the one of Annis (1999), invoking a *global* (in this instance, galactic) regulation mechanism, resetting local "astrobiological clocks" from time to time, and thus introducing correlation into *a priori* chaotic timescales of civilization development. This approach can be further developed and influence our practical considerations vis-à-vis feasibility of SETI projects, which is shown in Ćirković (2004). We have to keep in mind that, in most cases, *successful solution of Olum's problem will enable us to resolve Fermi's paradox as well.* That this is not exclusively so, testifies Olum's solution 2.9 ("large civilizations may consist of a small number of individuals"), which does not help us with Fermi's paradox; but this is an exceptional case.

Fermi's paradox can be usefully regarded as a boundary value problem. Physical theories (with partial exception of theories in quantum cosmology) are conventionally represented as a set of *dynamical equations* describing lawful behaviour of physical systems, usually matter fields. This is, of course, not enough, since in order to solve specific problems and to be able, consequently, to test the theory against its rivals, we need a set of boundary conditions for each specific problem. These can be given in a plethora of different ways, but if we wish to investigate evolution of a physical system in time, the natural thing to do is to specify *initial* or *final* conditions. We have a similar situation in astrobiology, *mutatis mutandis*: we may regard various ways of origin and evolution of life and intelligence as subject to (yet unknown) laws of *astrobiological dynamics*, but we still need specific boundary conditions in order to solve the particular problem of evolution of extraterrestrial life and intelligence in our Galaxy. Fermi's paradox tells us that *whatever the true astrobiological dynamics is, it must be consistent with existence of a large region, at present epoch, not filled with intelligent life*. How big that region is depends on how confident we are in detection capacities of our instruments, but it does not influence the general methodological point. Similarly, Olum's problem refers to the boundary conditions in probabilistic sense (*it is not overwhelmingly probable that random observers will belong to large civilizations at present*), and with additional cosmological input (spatially infinite universe).

<u>(II) Finitism.</u> The strategy which we shall *not* pursue here, but which could present a valid methodological point to be made, is that of *finitism*. If we accept, in accordance with Aristotle, that infinities exist only in potential, the limiting process for the fraction of observers in small civilizations $\lim_{V \to \infty} N_s \langle n_s \rangle / (N_l \langle n_l \rangle + N_s \langle n_s \rangle)$, is illegitimate. Here $N_s(V)$ and $N_l(V)$ are numbers of small and large civilizations in the volume *V* investigated, and $\langle n_s \rangle$, $\langle n_l \rangle$ average numbers of observers per small and large civilization, respectively. Instead of having an improbable value of $10^{-8}$, as Olum asserts, it would be undefined in a similar manner as the infinite series $\sum_{k=0}^{\infty} (-1)^k = 1 - 1 + 1 - 1 + ...$ has no convergent sum. This finitist view has been defended in more recent times by Dummett among philosophers and Brouwer and other intuitionists among mathematicians. Physical cut-off could be provided, as Olum mentions, by the existence of cosmological constant which would divide the spacetime into causally unconnected parts separated by event horizons. We shall not pursue this strategy here, since it seems less interesting and appealing than the evolutionary solution considered below.

(III) What time is it? "The problem will exist even if we confine ourselves to those observers who exist presently." This reasonable stance implies that we have *indexical knowledge* on the epoch we are living in. This is an important piece of additional data which has to be taken into account in the anthropic reasoning. The quoted sentence of Olum, for instance, could not be uttered if "presently" were to apply to epochs incompatible with the existence of intelligent observers, e.g., the times before galaxy formation, or the epoch in our distant future when there is going to be no negentropy sources left. Although it sounds quite tame, this simple constraint may, in fact, subtly undermine any reasoning, tacitly assuming our understanding of the necessary conditions for the existence of observers and civilizations.

## 3. A new solution: taking evolution into account

The devil hides in the details. When Olum writes: "Something must be wrong with our understanding of how civilizations *evolve* if only one in a billion can survive to colonize its galaxy," (Sec. 2.5, emphasis added), he is on the right track. Unfortunately, he doesn't offer a glimpse of what that "wrong" might be. Let us try to fill in the gap here.

The title of the relevant section of Olum's paper is "Infinitesimally few civilizations become large". This is an instance of sliding into the Parmenidian timeless view so dear to philosophers.[1] The correct title would be "Infinitesimally few civilizations have become large so far". There is no inconsistency here. The universe, be it infinite or finite, evolves: it changes with cosmic time.[2] Platonism has been largely excised from modern physical cosmology in mid-1960s after the resolution of the great "cosmological controversy" (Kragh 1996) between conflicting paradigms of the Big Bang and the steady-state cosmologies. What has been a sufficient condition for X at epoch $t_1$, is not necessarily sufficient at the epoch $t_2$. This is so, on the grounds of principle, but in this case we can do even better in specifying changes relevant to our topic.

---

[1] Or, in a less merciful Nietzschean term, "mummificationist" view. In a beautiful passage in *Twilight of the Idols*, philosophers are rebuked for "their lack of historical sense, their hatred of even the idea of becoming… They think they are doing a thing *honor* when they dehistoricize it, *sub specie aeternitatis* – when they make a mummy of it." (Nietzsche 1968, p. 351).

[2] There is a set of necessary conditions for the universal "cosmic time" to be well-defined. They are satisfied in all currently favoured cosmological models, and in all inflationary models Olum considers. They are not satisfied, for instance, in the celebrated rotating universe of Kurt Gödel (e.g., Stein 1970), which contains closed time-like curves.

There are strong empirical reasons to conclude that the universe has been less hospitable to life earlier in its history. One instance of such behaviour is related to the chemical enrichment of matter: fewer elements heavier than helium mean a smaller probability for the formation of terrestrial planets, and perhaps a smaller probability of biochemical processes leading to life, intelligence and observers. This has been recently spectacularly quantified by Lineweaver and collaborators who showed that, for instance, the median age of Earth-like planets in the Milky Way is 6.4 ± 0.9 billion years; before then conditions were far less favourable for the formation of possible life-bearing sites (Lineweaver et al. 2004). Another, possibly crucial effect, are catastrophes capable of disrupting the evolutionary sequence leading from the simplest prokaryotes to complex life, to animals, intelligent beings, and to civilizations, small and large. Notably, it has been recently confirmed that gamma-ray bursts detected by satellites in Earth's orbit at a rate of about one per day, are traces of explosions occurring in *any* galaxy within our cosmological horizon (for a review, see e.g., Mészáros 2002). Thus, their occurrence in any one particular galaxy (say ours), means a catastrophic event capable of destroying life forms in a large part or in the entire galactic habitable zone (Scalo and Wheeler 2002; Gonzalez, Brownlee, and Ward 2001). Fortunately enough, we also know from the cosmological research that their frequency *decreases with cosmic time*. Thus, the universe becomes more hospitable to life as the time passes.[3]

Actually, this seems to be true, for reasons unclear at present, for the Earth biosphere as well: Kitchell and Pena (1984) conclude that the extinction risk for species has been decreasing with time during the Phanerozoic eon. (Of course, this neglects the very recent human activities leading, unfortunately, to the "sixth" mass extinction; e.g., Wilson 2003.) Let us call the cumulative effect of all these occurrences the *hostility parameter*. The hostility parameter obviously has its spatial and temporal distribution; and these act as Bayesian constraints on any *a posteriori* estimates of our probability of finding ourselves in a large or small civilization.

Here, as in many other instances, we perceive the insufficiency of *gradualism*—an obsolete XIX century doctrine suggesting that no process occurring in the past was either qualitatively or quantitatively different from its present-day state (cf. Gould 1987). This is the same unsupported prejudice which impeded—and in some circles continues to impede—our progress in understanding the great mass extinctions of species in Earth's past (Raup 1991, 1994; Ward and Brownlee 2000). The sooner we discard it in the field of anthropic reasoning, as well as in astrobiology, the better off we are.

Let us take a look at a concrete simple ("toy") model which provides an illustration of this point. Observe a representative volume of space and count $N_0$ habitable sites where life and intelligence can develop rapidly.[4] Let's assume that the probability of a small civilization becoming a large one, evolves over time as:

$$p_l(t) = p_{l0}[1 - \exp(-t/\tau)], \qquad (1)$$

where $t$ is the cosmic time (measured from some relevant moment, say the galaxy formation), $\tau$ is the hostility parameter expressed as a characteristic timescale (say for gamma-ray burst rarefaction), and $p_{l0}$ is the asymptotic "standard" probability—*ceteris paribus*—of a small civilization making the transition to a large one. In Olum's study, this probability is at least 0.1 (since his "timeless" argument assumes that 90% of the currently existing civilizations are small ones).

In this toy model, the fraction of observers living in a large civilization is, clearly

$$f_{\text{large}}(t) = \frac{\text{number of observers in large civs}}{\text{total number of observers}} =$$
$$= \frac{\text{number of observers in large civs}}{\text{number of observers in large civs} + \text{number of observers in small civs}} = \qquad (2)$$
$$= \frac{N_0 p_l(t)\langle n_l \rangle}{N_0 p_l(t)\langle n_l \rangle + N_0 [1 - p_l(t)]\langle n_s \rangle}.$$

Here $\langle n_l \rangle = 10^{19}$ is the average number of observers in a large civilization, and $\langle n_s \rangle = 10^{10}$ is the average number of observes in a small civilization (using the very same numbers as in Olum's study). All numbers of observers are taken to be time-dependent, evolving quantities. Now, from (1) and (2), after trivial algebraic transformations, we obtain

$$f_{large}(t) = \frac{p_{l0}\langle n_l \rangle}{p_{l0}\langle n_l \rangle + \left[\frac{1}{1-\exp(-t/\tau)} - p_{l0}\right]\langle n_s \rangle} \approx \frac{\langle n_l \rangle}{\langle n_l \rangle + \frac{1}{p_{l0}}\frac{1}{1-\exp(-t/\tau)}\langle n_s \rangle}. \quad (3)$$

---

[3] Obviously, this tendency will reverse in far future of the universe, with the inexorable rise of entropy leading to the state close to the classical "heat death" (e.g., Adams and Laughlin 1997). But the timescale for this reversal is many orders of magnitude larger than the timescales of interest here.
[4] This is unrealistic, but is more in agreement with the Olum's scenario. We shall see below how this assumption may also be relativized with respect to the evolution of the universe in time.

(We neglect here the $p_{l0}\langle n_s \rangle$ term as arguably insignificant.) Upon inspection of the last expression in (3) we conclude that *at least one* of the following propositions must be true:

(1) $f_{large}(t) \approx 1$;

(2) $p_{l0} \approx 0$; or

(3) $\exp(-t/\tau) \approx 1$.

The proposition (1) is what Olum finds paradoxical (and corresponds perhaps to his solution 2.8, claiming that we are in fact part of a large civilization without been aware of that fact[5]). The proposition (2) corresponds to the idea that infinitesimally few civilizations *ever* become large ones, presumably because of some inherent problem like self-destruction. This might be valid for some small civilizations, but does not satisfy the non-exclusivity requirement, as both Olum and many previous investigators of Fermi's paradox concluded. But (3) is something entirely new. Here we have a *global evolutionary effect* acting to impede the formation of large civilizations. *It does not clash with any observation, as (1) does, nor does it imply something about (arguably nebulous) alien sociology, as does (2)*.

Of course, our toy model is certainly not realistic. One way of improving it would be to perceive that there might be many "critical steps" (Carter 1983; Maynard Smith and Szathmary 1997; Knoll and Bambach 2000) in evolving toward a large civilization. Each critical step takes some time. For instance, we may envisage the transition from simple life to complex life as a step having its own hostility timescale (suppressing and postponing "Cambrian explosion analogs" everywhere), and the same for transition between complex life and intelligence, intelligence and small technological civilization, etc. In the simplest analogy with the toy model above, we could substitute a single term in the denominator of (3) with something like

$$\prod_{i=1}^{n} \frac{1}{1-\exp(-t/\tau_i)}. \qquad (4)$$

---

[5] Of course, some very weak form of the principle of indifference is necessary here for the said conclusion. Nevertheless, this option is quite implausible on other grounds, as Olum points out.

Clearly, in this case, we would not need to worry about Olum's problem as long as the proposition *p*

$$p: (\exists i)\ \tau_i \gg t \qquad (5)$$

is true. It is obvious that *p* is not true for all times, so at some particular epoch we would have to face the problem again. Fortunately enough, that epoch lies in the distant cosmological future, so we need not worry about it now.

**4. Discussion: Too Early to Worry?**

By taking into account the physical evolution of the universe and the underlying requirements for civilizations (either large or small), we can help resolve the problem, pointed out by Olum, in a manner quite in accordance with current tenets of empirical astrobiology and anthropic reasoning. In the spirit of Olum's section/solution 2.10 ("Many factors acting together"), we may conclude that global evolutionary effects, in any case, constitute a large piece of the ultimate explanation, while local factors, like the loss of individuality, would account for the rest. Parenthetically, this argues against mildly sensationalistic title of Olum's paper ("Conflict between anthropic reasoning and observation"), since our present observation, incomplete as it is in the astrobiological field, should not yet be contraposed with the anthropic reasoning.[6]

The solution presented here should not be misconstrued as simplistic assertion that seemingly paradoxical situation *has not occured yet*, in the same sense as a lottery player can assert that a particular number of his has not come up yet. Current astrobiological state-of-affairs of our Galaxy is not a sort of a fortuitous state-of-affairs randomly pulled from the lottery bowl. Contrariwise, our intention is to emphasize systematic and on the average deterministic *and* predictable evolutionary processes in the physical reality. Such processes have precluded (in the probabilistic sense) "paradoxical" states-of-affairs from arising so far. Qualification "on the average" is necessary here, since although astronomers cannot, for example, predict exactly where and when will a particular supernova or gamma-ray burst occur, it is fairly incontroversial that the general trend of decreasing frequencies can be inferred on the basis of large observational surveys.

As we have seen, with increasing time elapsed from the galaxy formation epoch, the chance of finding a large civilization gradually increases. In that sense—which is only a trivial application of gradualism—we can assert that the universe becomes more hospitable to life and intelligence. In the same manner, the problem of conflict between anthropic reasoning and observations will become more acute—under the assumption that *both* anthropic reasoning *and* our observations do not change in future. However, it is reasonable to conjecture that our observations regarding this issue are going to change, by either discovery of a large civilization in our past light cone, or by *becoming* a large civilization ourselves. However, even under the most optimistic timescales, this is not a very pressing concern. (And it can be argued, as well, that the rapid development of astrobiology will, by that time, resolve the problem on empirical grounds.)

We conclude that it is *too early* (on the cosmological timescale) in the history of the universe for a situation to arise which contains Olum's problem, and that—given the present situation—it is *too early* (on human timescale and in the slightly different sense of being premature) to conclude that either astrobiology or our understanding of the observation selection conflicts with observation.

**Acknowledgements**. Insightful comments of a referee for *Biology and Philosophy* enormously helped in improving a previous version of this manuscript. It is a pleasure to acknowledge Oxford Colleges Hospitality Scheme and Dr. Nick Bostrom for a pleasant time during which this paper has been conceived. I use this opportunity to express my gratitude to Ken D. Olum, Istvan Aranyosi, Karl Schroeder, Cosma R. Shalizi and Slobodan Popović for useful comments and suggestions, as well as to Vesna Milošević-Zdjelar and Alan Robertson for invaluable technical help. This is an opportunity to thank *KoBSON* Consortium of Serbian libraries, which enabled at least partial overcoming of the gap in obtaining the scientific literature during the tragic 1990s.

---

[6] At least this applies to serious forms of anthropic reasoning, like those developed by Bostrom (2001, 2002), which demonstrate its disteleological nature.